\documentclass[11pt,a4paper]{article}
\usepackage{jheppub}
\usepackage{subfig}

\newcommand{\ba}{\begin{eqnarray}}
\newcommand{\ea}{\end{eqnarray}}

\title{Cold Electroweak Baryogenesis in the\\ Two Higgs-Doublet Model}

\author[a]{Anders Tranberg,}
\author[b]{Bin Wu}

\affiliation[a]{Niels Bohr International Academy and Discovery Center, Niels Bohr Institute,\\  Blegdamsvej 17, 2100 Copenhagen, Denmark}
\affiliation[b]{Faculty of Physics, University of Bielefeld, Bielefeld, Germany}

\emailAdd{anders.tranberg@nbi.dk}
\emailAdd{binwu@physik.uni-bielefeld.de}

\abstract{We perform the first investigation of cold electroweak baryogenesis in the two Higgs-doublet model (2HDM). The electroweak symmetry breaking transition is assumed to occur through a spinodal instability from a super-cooled initial state. We consider the creation of net Chern-Simons number, which through the axial anomaly is equivalent to a baryon asymmetry. CP violation is explicit in the scalar potential, but only in combination with P-violation is it possible for an asymmetry to be generated. This is introduced through the leading C-and P-breaking, but CP invariant, term expected to arise upon integrating out the fermions in the theory. We perform real-time lattice simulations of the transition, and find the coefficient of this term required for successful baryogenesis.}


\begin{document}

\maketitle

\section{Introduction}

The origin of the baryon asymmetry in the Universe is a major unresolved issue in cosmology, and constitutes together with dark matter, cosmological inflation and dark energy the only direct evidence of physics beyond the Standard Model. Baryon number violating processes do exist within the electroweak sector of the Standard Model, and this realization has prompted extensive development of the Electroweak Baryogenesis scenario \cite{EWBG}. Within the Standard Model, one encounters two stumbling blocks to successful baryogenesis: CP-violation as encoded in the CKM quark mixing matrix is much too small at high temperatures ($T\sim m_{\rm w}$) \cite{shaposhCP} and the electroweak phase transition is too smooth to supply the out-of-equilibrium conditions necessary for an asymmetry to be created \cite{rummukainen}. 

Three main avenues are pursued to remedy these shortcomings. Firstly, a non-trivial mass sector for the leptons may provide for leptogenesis \cite{lepto1,lepto2}, with additional CP-violation in the lepton mass matrix and heavy masses allowing for out-of-equilibrium decay. Secondly, one may extend the Higgs sector to allow for a strongly first-order electroweak transition as well as CP-violation through complex couplings, reviving the original ``hot'' electroweak baryogenesis mechanism, see for instance \cite{fromme,konstandin}. And thirdly, additional scalars may allow for the Universe to super-cool below the electroweak scale, and then trigger a low-temperature spinodal transition. Over the last decade, this has become known as ``cold'' electroweak baryogenesis \cite{GB1,krausstrodden,copeland,CP3d}. 

The ``cold'' scenario was originally an attempt to bypass the lack of a strong equilibrium phase transition by invoking an out-of-equilibrium process such as preheating after inflation, and necessary CP-violation was introduced through a hypothetical higher order bosonic operator \cite{GB1,CP3d}. Through extensive numerical simulations, it was shown that this could lead to successful baryogenesis \cite{CP3d,finitekap,finitet}. 

Later, Standard Model CP-violation itself was reintroduced in this ``cold'' setting. As was argued in \cite{smit4,schmidt6,salcedo6,salcedo8}, and more explicitly computed recently \cite{dim81,dim82,CPV1}, the effect is not strongly suppressed at low temperatures. It was argued that if the transition has an effective temperature around 1 GeV, Standard Model CP-violation may be able to provide for baryogenesis. This awaits further study.

In the 2HDM, which is our main concern in the present work, we will neglect the CKM matrix. Instead, CP-violation is present in the scalar interactions. In addition, the potential has to be such that the potential exhibits a strong spinodal instability (see below). We will assume that the Higgs fields are coupled to the Standard Model fermions, but include these only through what we expect to be the leading (C- and P-breaking) bosonic term arising from integrating them out of the path integral. The reason for this is that for bosonic simulations, the asymmetry is generated as non-zero average Chern-Simons number, which is both CP- and P-odd. The scalar potential breaks CP through breaking C, and so P is conserved. So we need the effect of the explicit P-breaking present in the electroweak gauge-fermion interactions, in our case through a bosonic C- and P- breaking operator\footnote{This is in fact very similar to the situation in the Standard Model, except that there both the CP-breaking and the C-/P-breaking are associated with the fermions.}.

Our goals are therefore threefold: To simulate a cold tachyonic symmetry breaking transition in the 2HDM. To study the dynamical interplay between a source of C-violation and one of C- and P-violation to generate effective P- and CP-violation, necessary for the creation of a baryon asymmetry. And finally to explicitly compute the magnitude of  the asymmetry as a function of in particular the coefficient of the C- and P-breaking term. 

\paragraph{Electroweak baryon number creation:}

In the electroweak sector of the Standard Model, an anomaly relates the fermion (baryon B, lepton L) numbers to the evolution of the Chern-Simons number $N_{\rm cs}$ of the SU(2) gauge field
\ba
\label{eq:anomaly}
B(t)-B(0)=L(t)-L(0)=3\left[N_{\rm cs}(t)-N_{\rm cs}(0)\right]=3\int dt\int d^3x\, \frac{1}{16\pi^2}\textrm{Tr}\, F_{\mu\nu} \tilde{F}^{\mu\nu},
\ea 
with $\tilde{F}^{\mu\nu}=1/2\epsilon^{\mu\nu\rho\sigma}F_{\rho\sigma}$. The question of baryogenesis then becomes a question of whether permanent change of Chern-Simons number can take place. It so happens that the vacuum structure of the theory is that of a sequence of degenerate minima corresponding to integer values of the Chern-Simons number. Since these vacua are ``pure gauge'', in the minima Chern-Simons number coincides with the topological winding number of the Higgs field; but whereas Higgs winding is always integer, Chern-Simons number can change continuously between vacua. 

Adjacent vacua are separated by potential barriers, and in equilibrium, the rate of transition from one to the other is a diffusive process governed by the Boltzmann factor of the saddle point configuration at the top of the effective potential, known as the Sphaleron \cite{thooft,manton}. The Sphaleron energy is in turn dominated by the need for the Higgs winding number to wind in such a transition, since this can only happen through a local (in space) zero of the Higgs field length. The Sphaleron rate is by now very well known numerically \cite{sphaleronrate,sphaleronrate2}. 

Out of equilibrium, on the other hand, there is no concept of temperature and no Boltzmann distribution. The dynamics of Chern-Simons number change need not be diffusive, and the change of Higgs winding number can happen spontaneously where the Higgs field happens to go through zero. This can happen through analogues of the Kibble mechanism \cite{krausstrodden,turok1,turok2}, through overall oscillations of the Higgs field \cite{GB1,CP3d,sexty} or through statistical fluctuations as sphaleron-like half-knots \cite{sexty}. In particular, Chern-Simons number and Higgs winding need not follow each other until the system begins to relax to equilibrium.

In the 2HDM, we now have two Higgs winding numbers (one for each doublet), both of which must agree with the Chern-Simons number in the vacuum. Therefore, both must change in order to generate baryons, and so both Higgs fields must locally go through zero length during the evolution. This need not happen at the same time, nor at the same time as Chern-Simons number transitions, as long as they all end up at the same integer value. All in all this provides for some highly non-trivial dynamics. 
 
\paragraph{Spinodal transition:}

The Higgs field potential is such that at low temperature the electroweak symmetry is spontaneously broken through the appearance of a Higgs expectation value. The assumption is that at high temperatures this symmetry is restored and the expectation value vanishes as the effective potential changes. In the Standard Model we know that the transition is a cross-over \cite{rummukainen}, and that the rate of change of temperature in the Universe is governed by the Hubble rate. Since at the electroweak scale $H\simeq 10^{-5}$ eV, the system will remain very close to equilibrium throughout.  

Extensions of the scalar sector of the theory allows for a strongly first order phase transition, in which case bubble nucleation takes place, a very much out of equilibrium process. Baryogenesis can then occur at the bubble wall as it sweeps through the plasma \cite{EWBGrev}. 

We will be interested in another possibility, namely where the tree level potential is such that electroweak symmetry breaking is triggered from a super-cooled state. Consider as a toy model the potential 
\ba
V(\phi,\sigma)=\left(g^2\sigma^2-\mu^2\right)\phi^\dagger\phi+\lambda(\phi^\dagger\phi)^2+V_0+V(\sigma),
\ea 
($V(\sigma)$ is some appropriate potential for $\sigma$ only, and $V_0$ fixes the minimum to have $V=0$.) As long as $\langle\sigma\rangle>\mu/g$, there is no symmetry breaking, and the Universe will cool down under expansion, in principle to arbitrarily low temperature. If the dynamics of $\sigma$ is then such that it goes from $\sigma >\mu/g$ to $\sigma \ll \mu/g$ very fast (on the time scale of $\mu$), the $\phi$ field will experience a fast mass quench and IR modes will become unstable. This can be seen from the linearized Klein-Gordon equations with negative mass-squared, in momentum space
\ba
\bigg[\partial_t^2+k^2-\mu^2\bigg]\phi_{\bf k}=0\rightarrow \phi_{\bf k} \propto a_{\bf k}e^{i\sqrt{\mu^2-k^2}t}+a_{\bf k}^\dagger e^{-i\sqrt{k^2-\mu^2}t},
\ea
so that for $|{\bf k}|<\mu$, occupation numbers grow exponentially
\ba
n_{\bf k}\propto e^{2\sqrt{\mu^2-k^2}t},
\ea
until nonlinear interactions kick in and curb the growth. Large occupation numbers lead to approximate classical behaviour, which is why it is appropriate to evolve the system classically. A more detailed exposition of this point can be found in \cite{GB3,CP1d}. Specific realizations of $V(\sigma)$ can be found in \cite{vantent,enqvist,servant}.

\paragraph{Classical Bosonic simulations:}

Very recently, an approach to simulate the whole SU(2)-Higgs-fermion system numerically has been developed \cite{hindmarshborsanyi,fermions1,fermions3}, but the numerical implementation of the quantum fermions is very challenging in terms of computer resources. So for broad numerical investigations, it is vastly more efficient to consider an effective bosonic treatment, where the effect of fermions is captured through a series of higher order bosonic operators.

As is always the case for effective bosonic simulations of electroweak baryogenesis, it is then assumed that when the gauge field Chern-Simons number changes in time, the baryon number of the fermions coupled to it follows along according to the anomaly equation (\ref{eq:anomaly}). Hence we are interested in numerically computing $\langle N_{\rm cs}(t)\rangle$, in our case using classical evolution equations for the fully coupled and non-perturbative SU(2)-2Higgs system, including CP-violation (from the tree-level scalar potential) and C- and P-violation (from effective higher order bosonic operators).

The numerical procedure is then that the full quantum average of a certain operator is replaced by an ensemble average over field realizations
\ba
\overline{\mathcal{O}}_{\rm quant.}\rightarrow \overline{\mathcal{O}}_{\rm clas. ens.},
\ea
and that each member of the ensemble is evolved independently using classical evolution equations. As mentioned, classical behaviour is applicable for fields with large occupation numbers as after a spinodal transition.

The paper is organized as follows: In section \ref{sec:model} we will set up the 2HDM and the choice of scalar potential we use. In section \ref{sec:init} we describe the numerical setup and the initial conditions. In section \ref{sec:fullnum} we present our numerical results, in \ref{sec:single} demonstrating why only CP-violation or only P- and C-violation does not generate an asymmetry and in \ref{sec:averages} performing simulations of a whole ensemble of configuration with both CP- and C- and P-breaking, computing the size of the asymmetry as a function of the strength of P-violation. We conclude in section \ref{sec:conclusion}. Some details of the choice of parameters for the Higgs potential can be found in appendix \ref{app:parameters}.

\section{The 2HDM}
\label{sec:model}

The 2HDM is defined through the continuum action
\ba
S=-\int d^3x\, dt\,\bigg[
\frac{1}{4g^2}\textrm{Tr}\, F_{\mu\nu}F^{\mu\nu}+
(D_\mu\phi_1)^\dagger D^\mu\phi_1+(D_\mu\phi_2)^\dagger D^\mu\phi_2+V(\phi_1,\phi_2)
+S_\textrm{C/P}
\bigg],\nonumber\\
\ea
where we use the metric $\eta={\rm diag}(-+++)$, $\phi_{1,2}$ are SU(2) doublets with hypercharge $+1$ and $F^{\mu\nu}$ is the field strength tensor of the gauge field. The covariant derivative is
$D_\mu\phi_i=(\partial_\mu+iA_\mu)\phi_i$ and the potential is in all generality
\ba
V(\phi_1,\phi_2)&=&-\mu_{11}^2\phi_1^\dagger\phi_1-\mu_{22}^2\phi_2^\dagger\phi_2-\mu^2_{12}\,\phi_1^\dagger\phi_2-\mu^{2,*}_{12}\phi_2^\dagger\phi_1\nonumber\\
&&+\frac{\lambda_1}{2}(\phi_1^\dagger\phi_1)^2+\frac{\lambda_2}{2}(\phi_2^\dagger\phi_2)^2+\lambda_3(\phi_1^\dagger\phi_1)(\phi_2^\dagger\phi_2)+\lambda_4(\phi_2^\dagger\phi_1)(\phi_1^\dagger\phi_2)\nonumber\\
&&+\frac{\lambda_5}{2}(\phi_1^\dagger\phi_2)^2+\frac{\lambda_5^*}{2}(\phi_2^\dagger\phi_1)^2
+\lambda_6(\phi_1^\dagger\phi_1)(\phi_1^{\dagger}\phi_2)
+\lambda_6^*(\phi_1^\dagger\phi_1)(\phi_2^{\dagger}\phi_1)\nonumber\\
&&+\lambda_7(\phi_2^\dagger\phi_2)(\phi_1^{\dagger}\phi_2)
+\lambda_7^*(\phi_2^\dagger\phi_2)(\phi_2^{\dagger}\phi_1).
\ea
The parameters $\lambda_{1,2,3,4}$ and $\mu^2_{11,22}$ are real and in general $\lambda_{5,6,7}$ and $\mu_{12}^2$ are complex. 

For the bosonic theory as written here, a generalized CP-symmetry can be defined which includes Higgs field basis transformations (in $\phi_{1,2}$-space) \cite{2hdmreview,CP2}. In this general sense, CP is then only violated if there is no such transformation that can make all the parameters real\footnote{We will not consider the possibility of spontaneous CP-symmetry breaking in this work.}. 

However, when the scalar fields are coupled to fermions in the usual way and the Yukawa couplings and CKM mixing matrix is fixed, there is no such freedom in general. In our case, we do not explicitly have fermions but imagine that they have been integrated out in the path integral, and that the leading effective bosonic operator resulting from this is \cite{turok1,turok2}\footnote{Turok and Zadrozny \cite{turok1,turok2} consider a linearized, gauge fixed version of this term, where $i(\phi_1^\dagger \phi_2-\phi_2^\dagger \phi_1)\rightarrow h_1h_2\theta$, with $\phi_1=(0,h_1e^{i\theta_1})^T$, $\phi_2=(0,h_2e^{i\theta_2})^T$ and $\theta=\theta_1-\theta_2$.}
\ba
\label{eq:CandP}
S_{\rm C/P}=\frac{\delta_{\rm C/P}}{16\pi^2 m_W^2}i(\phi_1^\dagger \phi_2-\phi_2^\dagger \phi_1)\textrm{ Tr}\,F_{\mu\nu}\tilde{F}^{\mu\nu},
\ea
where the Yukawa couplings and the mixing matrix is encoded in the real parameter $\delta_{\rm C/P}$. This term breaks C and P maximally, but conserves the combination CP. Many more operators appear upon integrating out the fermions, but we will neglect all the C-even, P-even ones since they do not participate in generating an asymmetry, and take (\ref{eq:CandP}) as representative of the potentially many C- and P-violating terms at leading order. A complete calculation can be performed along the lines of \cite{CPV1}, also providing a value for the coefficient $\delta_{\rm C/P}$. We again emphasize that we ignore the Standard Model CKM matrix as a source of CP violation. Had we included it, additional C-breaking, P-conserving (and hence CP-violating) terms would appear \cite{CPV1}, but at higher order in gradients. 

From now on, we will set $\lambda_{6,7}=0$ for simplicity and we will fix the rest of the parameters in the following way (see also appendix \ref{app:parameters}):
\begin{itemize}
\item The five Higgs particle modes have masses $m_{\pm}=400$ GeV, $m_1=125$ GeV, $m_2=300$ GeV, $m_3=350$ GeV. The lightest mode represents the Standard Model-like Higgs \cite{LHC1,LHC2}, and the remaining masses are chosen large enough to have avoided detection, while being ``generic'' in the sense of non-degenerate and not having some accidental symmetry.
\item The Higgs vevs are set by $|v_1|/|v_2|=\tan\beta=2$ and $v_1^2+v_2^2=246^2$ GeV$^2$.
\item The electroweak transition is driven by the spinodal instability of one or both of the Higgs fields. We will consider only the case when both of the eigenvalues of the matrix
\ba
\label{eq:massmatrix}
M^2=\left(\begin{array}{cc}
-\mu_{11}^2&-\mu_{12}^2\\
-\mu_{12}^{2,*}&-\mu_{22}^2
\end{array}\right)
\ea
are negative. We indeed found that having both fields unstable (rather than only one) makes the system evolve faster and that the asymmetry becomes larger.
\item CP-conserving, parameter set 1: If all the parameters are chosen real, there is no CP-violation, and the vevs are real\footnote{See footnote 2.}. This leaves two free parameters.
\item Real vev, parameter set 2: We can have CP-violating also keeping both vevs real. In this case $\lambda_5$ is complex, but obeys the constraint $\textrm{Im}(\lambda_5)v_1v_2=\textrm{Im}(\mu_{12}^2)$. We are left with three free parameters.
\item Complex vev, parameter set 3: Finally, we can make Higgs basis transformations and make $\lambda_5$ real, and the $\phi_2$-vev (for instance) complex. $\mu_{12}^2$ is then also transformed. We again have three free parameters, and the potential is the same as for the real-vev case, but centered around the complex minimum. Then keeping $\delta_{\rm C/P}$ unchanged (the coupling to fermions), it is a different system to parameter set 2.
\end{itemize}
We note that it really is C that is broken by complex parameters, whereas P is always conserved by the scalar potential.

\section{Numerical setup and initial conditions}
\label{sec:init}

The continuum theory is discretized on a space-time lattice of volume $L^3=a_x^3n_x^3$ in a similar way to \cite{CP3d}. The lattice spacing is set through $a_x v=0.6$, where $v=246\,$GeV, and the lattice size is $Lv=n_x a_xv=38.4$, i.e. $n_x=64$. In terms of the smallest and largest mass scales present, we have $a_xm_1\simeq 0.3$, $am_{\pm}\simeq 0.98$ and $Lm_1\simeq 19.2$, $Lm_{\pm}\simeq 62.4$ which we consider acceptable for the level of precision aimed at here\footnote{For classical simulations there is no need to take a strict continuum limit, as long as results vary little with lattice spacing and volume.}. 

The classical equations of motion are derived by straightforward variation of the action, and we consider the following observables:
\begin{itemize}
\item The Higgs field expectation values, normalized to their vacuum values
\ba
\label{eq:higgsexp}
\bar{\phi}_1^2=\frac{2}{L^3v_1^2}\sum_xa_x^3\, \phi_1^\dagger\phi_1(x),\qquad \bar{\phi}_2^2=\frac{2}{L^3v_2^2}\sum_xa_x^3\, \phi_2^\dagger\phi_2(x),
\ea
\item We also need a measure for the angle between the fields, which encodes the C(P) violation in time, as well as being the first factor of (\ref{eq:CandP})
\ba
\label{eq:higgsangle}
\textrm{Im}(\phi_{2}^*\phi_{1})=\frac{1}{L^3v_1v_2}\sum_xa_x^3\,i\left(\phi_1^\dagger\phi_2-\phi_2^\dagger\phi_1\right)(x).
\ea
In addition to the average, we will consider the distribution of the local quantity inside the sum. 
\item The Chern-Simons number
\ba
\label{eq:ncs}
N_{\rm cs}(t)-N_{\rm cs}(0)=\int dt\, \int d^3x \frac{1}{16 \pi^2}\textrm{Tr}\,F_{\mu\nu}\tilde{F}^{\mu\nu}.
\ea
\item The Higgs winding numbers
\ba
\label{eq:higgswinding}
N_{\rm w}^a=-\frac{1}{24\pi^2}\int d^3x\,\epsilon_{ijk}\textrm{Tr}\left[U_a^\dagger\partial_i U_aU_a^\dagger\partial_j U_aU_a^\dagger\partial_k U_a\right],\quad a=1,2,
\ea
where 
\ba
\Phi_{a}=\left(
\phi_{a}, i\sigma_2\phi_a^*\right),\qquad U_a=\frac{1}{|\Phi_a|}\Phi_a.
\ea
The lattice winding numbers are integer up to discretization errors and both coincide with the Chern-Simons number in the vacua.
\item As a check on the numerical code, we keep track of Gauss constraint and energy conservation. 
\end{itemize}

We generate a set of random initial field configurations in the following way: Given the masses and couplings, we diagonalize the Higgs mass matrix (\ref{eq:massmatrix}), to find
\ba
\label{eq:massmatrix2}
M^2_d=\left(\begin{array}{cc}
M_{A}^2&0\\
0&M_{B}^2
\end{array}\right),
\ea
in terms of the eigenbasis $\phi_{A,B}$. We restrict ourselves to cases where $M_{A}^2$ and $M_{B}^2$ are both negative, and so both eigenmodes will have spinodally unstable (lattice) momentum modes, given by\footnote{If only one of the eigenvalues $M_{A,B}^2$ is negative, only the unstable modes of the corresponding field should be initialized.}. 
\ba
k^2_{\rm lat}=\sum_{i}\frac{1}{a_x^2}\left[2-2\cos\left(\frac{2\pi}{L}n_i\right)\right]<|M^2_{A,B}|,\quad i=1,2,3,\quad n_i=0,...,n_x-1.
\ea
These (and only these) momentum modes are then initialized in the ``quantum vacuum'', with a Gaussian distribution according to \cite{GB3,CP1d}
\ba
\langle\phi_{A,B}(k)\,\phi_{A,B}^\dagger(k)\rangle=\frac{1}{2}\frac{1}{\sqrt{k^2_{\rm lat}+|M_{A,B}^2|}},\qquad \langle\partial_t\phi_{A,B}(k)\,\partial_t\phi_{A,B}^\dagger(k)\rangle=\frac{1}{2}\sqrt{k^2_{\rm lat}+|M_{A,B}^2|}.\nonumber
\ea
This mimics a fast mass quench, where the potential instantaneously flips as
\ba
t<0,&\qquad& V(\phi_A,\phi_B)= (\phi_A^\dagger,\phi_B^\dagger)\left(\begin{array}{cc}
|M_{A}^2|&0\\
0&|M_{B}^2|
\end{array}\right)
\left(\begin{array}{c}\phi_A\\\phi_B\end{array}\right),\\
t>0,&\qquad& V(\phi_A,\phi_B)= (\phi_A^\dagger,\phi_B^\dagger)\left(\begin{array}{cc}
M_{A}^2&0\\
0&M_{B}^2
\end{array}\right)
\left(\begin{array}{c}\phi_A\\\phi_B\end{array}\right)+\textrm{quartic}.
\ea
The initialized fields are then rotated back to the original $\phi_{1,2}$ basis.

This is completely analogous to the approach in \cite{CP3d}, generalized to more than one Higgs field. Similarly, the gauge field is initialized $A_i=0$, and the simulations are done in temporal gauge $A_0=0$. The gauge momenta $\partial_t A_i$ are initialized by solving Gauss constraint in the background of the random Higgs field. 

Since we are interested in C-, P- and CP-violation, and in order to minimize the statistical noise, we generate an ensemble of initial configuration that is C-even and P-even. This is of course automatically satisfied for an infinitely large ensemble, since a configuration and its C- and/or P- conjugate have the same probability to be generated. But we make it explicit by, for all random configurations $\phi_{\rm init}({\bf x})$, $A_{\rm init}({\bf x})$ (in terms of link variables $U_i({\bf x})$), also including their C-, P- and CP- conjugate configurations in the ensemble, according to
\ba
&&\Phi_{i}\left( t, \vec{x} \right) \xrightarrow{P}\Phi_{i}\left( t, -\vec{x} \right),~~U_n\left( t, \vec{x} \right) \xrightarrow{P} U_n^\dagger \left( t, -\vec{x}- \hat{n} \right), \label{equ:parity}
\ea
and 
\ba
\Phi_{i}\left( t, \vec{x} \right) \xrightarrow{C}\Phi_{i}^*\left( t, \vec{x} \right),~~U_{n} \xrightarrow{C}U_n^*\left( t, \vec{x} \right).\label{equ:chargeConjugate}
\ea
In the continuum limit, this reduces to
\ba
\Phi_{i}\left( t, \vec{x} \right) \xrightarrow{P}\Phi_{i}\left( t, -\vec{x} \right),~~\vec{A}\left( t, \vec{x} \right) \xrightarrow{P} -\vec{A}\left( t, -\vec{x} \right),
\ea
and
\ba  
\Phi_{i}\left( t, \vec{x} \right) \xrightarrow{C}\Phi_{i}^*\left( t, \vec{x} \right), A_\mu\left( t, \vec{x} \right) \xrightarrow{C} - A^*_\mu\left( t, \vec{x} \right).
\ea
CP-conjugation is of course the combination of the two. We then construct averages for a given observable $\mathcal{O}$ through
\ba
\bar{\mathcal{O}}=\frac{1}{N_{\rm init}}\sum_{i} \mathcal{O}(\{\phi_i\}),
\ea
where
\ba
\mathcal{O}(\{\phi_i\})=\frac{1}{4}\left[\mathcal{O}(\phi_i)+\mathcal{O}(\phi^{C}_i)+\mathcal{O}(\phi^P_i)+\mathcal{O}(\phi^{CP}_i)\right].
\ea
This is identical to a normal average over all 4$\times N_{\rm init}$ configurations. But for the statistical error we use
\ba
\label{eq:sigma}
\sigma^2_{\mathcal{O}}=\frac{\overline{\mathcal{O}^2(\{\phi\})}-\left(\overline{\mathcal{O}(\{\phi\})}\right)^2}{N_{\rm init}}.
\ea

We can convert the non-zero average Chern-Simons number to a baryon asymmetry using the conversion prescription (we use the Standard Model content of fermionic and bosonic degrees of freedom),
\ba
\frac{n_B}{n_\gamma}=\frac{3\overline{N_{\rm cs}}/L^3}{(2\pi^2/45 g^* T^3)/7.04},\quad \frac{\pi^2}{30}g^* T^4 = V(0,0)-V(v_1,v_2),
\ea
i.e. distributing the initial potential energy on a thermal ensemble of all the relativistic degrees of freedom $g^*$. For the Standard Model degrees of freedom with masses less than $m_{\rm w}$, we have $g^*=86.25$. This amounts to
\ba
\label{eq:fin_asym}
\frac{n_B}{n_\gamma}=\overline{N_{\rm cs}}\times 1.2\times 10^{-4}\times \left(\frac{V(0,0)-V(v_1,v_2)}{v^4}\right)^{-3/4}.
\ea
The reheating temperature is, for the three choices of potentials respectively, 
\ba
\frac{T}{\textrm{GeV}}\simeq\left(99,104,104\right).
\ea
This is much higher than in the Standard Model, simply because the potential is deeper (we may even consider treating the W$^\pm$ and Z as relativistic, changing the temperature by a factor $(95.25/86.25)^{1/4}=1.025$, i.e. $2-3$ GeV lower temperature). Consequently, the thermal electroweak transition is also at higher temperature than in the Standard Model, and the sphaleron rate is modified accordingly. A detailed study of these features of the 2HDM model would certainly be of great interest both for hot and cold electroweak baryogenesis. 

There is an interesting twist, in that for the simulations performed here, the total number of degrees of freedom participating in the dynamics is $g^*=14$ (or $g^*=10$, if we neglect the heavy Higgs modes), in which case the effective temperature goes up by a factor $(86.25/14)^{1/4}\simeq 1.6$, which brings it above the lowest Higgs mass of $125$ GeV. This means that the temperature is large enough to partially ``restore'' symmetry in one direction in field space. This is an (albeit physical) artefact of not including all the existing degrees of freedom in the dynamics. We will comment further on this below.

\section{Results: }
\label{sec:fullnum}

\subsection{Single trajectories and symmetries}
\label{sec:single}

\begin{figure}
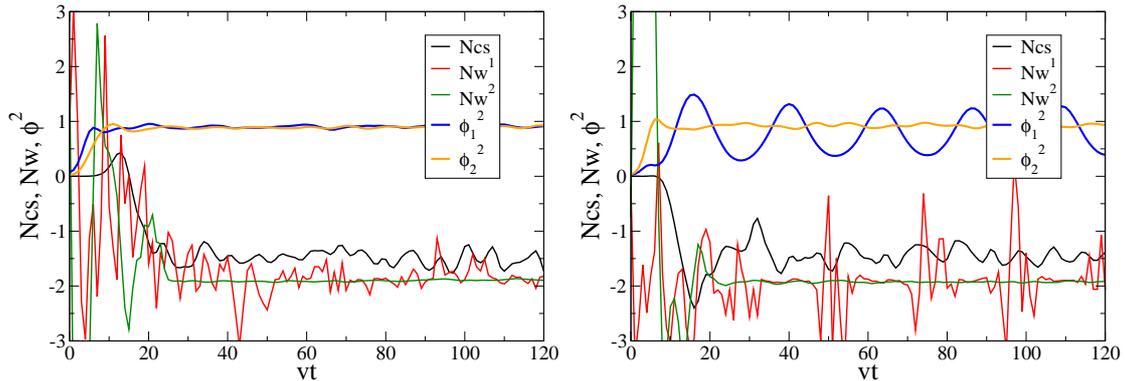

  \begin{center}
\epsfig{file=./Pictures/AllObsPot1.eps,width=0.48\textwidth,clip}
\epsfig{file=./Pictures/AllObsPot2.eps,width=0.48\textwidth,clip}
\caption{\label{fig:single} The Chern-Simons numbers, winding numbers and Higgs expectation values in time for a single configuration. Left: CP-conserving potential. Right: CP-breaking.}
\end{center}
\end{figure}

We will start out by considering a single initial field realization, and in Fig.~\ref{fig:single} (left), we show the Chern-Simons number Eq.~(\ref{eq:ncs}), both Higgs winding numbers Eq.~(\ref{eq:higgswinding}) and both Higgs expectation values Eq.~(\ref{eq:higgsexp}). We use the CP-conserving potential (parameter set 1). We observe that both Higgs fields perform a spinodal transition in a time $t\simeq 10/v$, ending up close to their respective vevs (here normalized to 1). Chern-Simons number oscillates and drifts in to $N_{\rm cs}\simeq -2$, and around $t=20/v$, both Higgs winding numbers settle close to Chern-Simons number. We also see that $N_{\rm w}^1$ is still quite noisy at later times, while $N_{\rm w}^2$ stays flat. This signals that the mass in the ``1'' direction is smaller than in the ``2'' direction, and that it is therefore easier for the field to make excursions near $\phi_1(x)\simeq 0$, where the lattice winding numbers are ill defined and easily ``flips''. The winding numbers a slightly off integer values because of discretization errors; the Chern-Simon number is not exactly integer because the fields have finite energy density/temperature.

In the right-hand figure we show the same observables, but for another single configuration evolving in a CP-breaking potential (parameter set 2). There is now a substantial difference in that the ``1'' mode is much lighter than the ``2'' mode. The ``2'' mode is strongly forced into its minimum, whereas the ``1'' field has large amplitude oscillations. The corresponding winding numbers are also different, in that $N_{\rm w}^2$ is smooth and closely integer, whereas $N_{\rm w}^1$ is noisy and has spikes and nearly jumps. This is because the field oscillation brings about zeros of $\phi_1(x)$. The spikes are also strongly correlated with the overall oscillations of the field. 

\begin{figure}
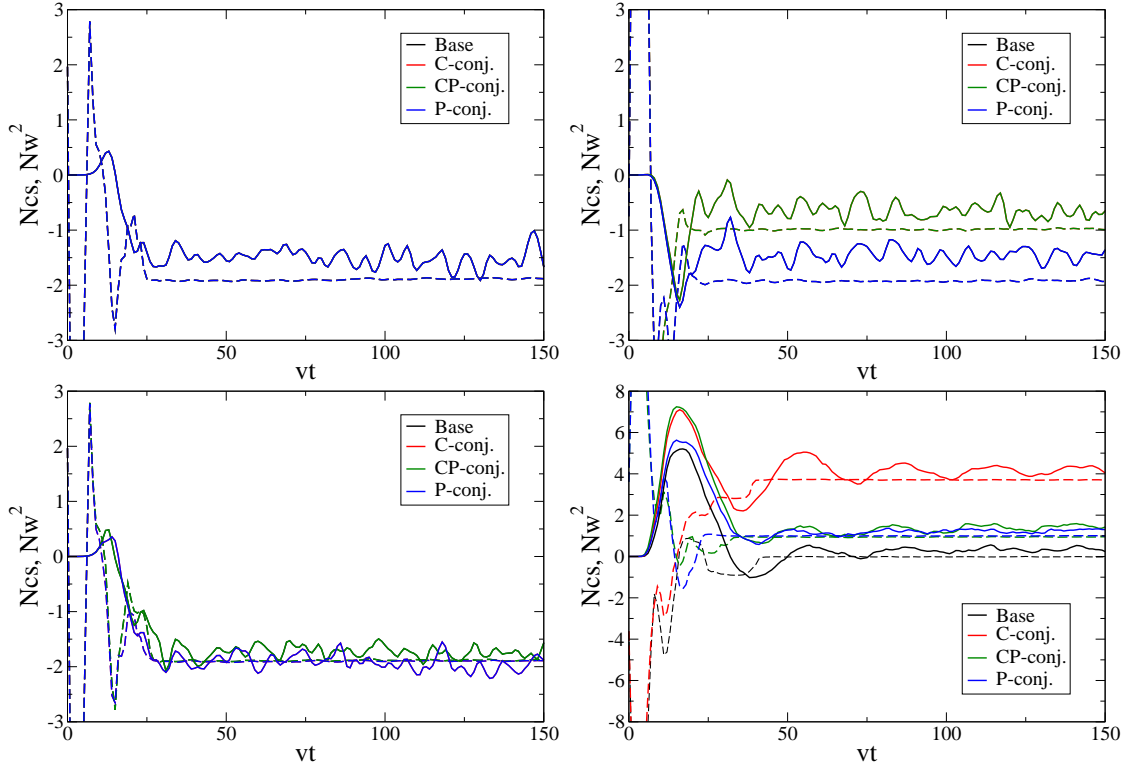

  \begin{center}
\epsfig{file=./Pictures/CPcons4.eps,width=0.48\textwidth,clip}
\epsfig{file=./Pictures/CPbreak4.eps,width=0.48\textwidth,clip}
\epsfig{file=./Pictures/Pbreak4.eps,width=0.48\textwidth,clip}
\epsfig{file=./Pictures/CPandPbreak.eps,width=0.48\textwidth,clip}
  \caption{\label{fig:single_all} The Chern-Simons number and winding number ``2'' for a single configuration and its C-, P- and CP- conjugate in the case of no C-, P- or CP-violation (top left) and when including CP-violation in the scalar potential (top right). Also the case of C-, and P-violation but no CP-violation (bottom left) and when including CP-violation as well (bottom right). The P- and CP- conjugates have been given an overall sign flip for illustration.}
\end{center}
\end{figure}

In Fig.~\ref{fig:single_all} we show the Chern-Simons number and winding number ``2'' for four conjugate initial configurations (base, its C-, P- and CP-conjugate), the P- and CP-conjugates with an overall sign-flip. In the top left plot, the potential conserves CP and the C-/P- breaking term is turned off $\delta_{\rm C/P}=0$. The curves are identical, and we conclude that the average Chern-Simons number is identically zero, since the P- and CP-conjugate exactly cancel the other two. In other words, $N_{\rm cs}(\{\phi\})=N^a_{\rm w}(\{\phi\})=0$. 

In the top right plot we show the Chern-Simons and winding numbers in a similar run, but now in a CP-breaking, real-vev potential (parameter set 2). The four configurations are no longer identical up to the overall sign, but pair up in P-conjugates. The base configuration cancels with its P-conjugate; the C-conjugate pairs up with the CP-conjugate. Again the averages $N_{\rm cs}(\{\phi\})$ and $N^a_{\rm w}(\{\phi\})$ are zero. 

Then in the bottom left plot we have included the P- and C-violating term Eq.~(\ref{eq:CandP}), but the potential is again CP-conserving. Now the pairing is the other way around: The base configuration is cancelled by the CP-conjugate, whereas the C- and P-conjugates average out. $N_{\rm cs}(\{\phi\})=N^a_{\rm w}(\{\phi\})=0$. Finally, including both the CP-breaking potential and the C-/P-breaking term, we find as shown in the bottom right plot that the four trajectories of Chern-Simons number are all different. $N_{\rm cs}(\{\phi\})\neq 0$ and $N_{\rm w}(\{\phi\})\neq 0$. Hence a net baryon asymmetry is created.

\subsection{Ensemble averages}
\label{sec:averages}

\begin{figure}
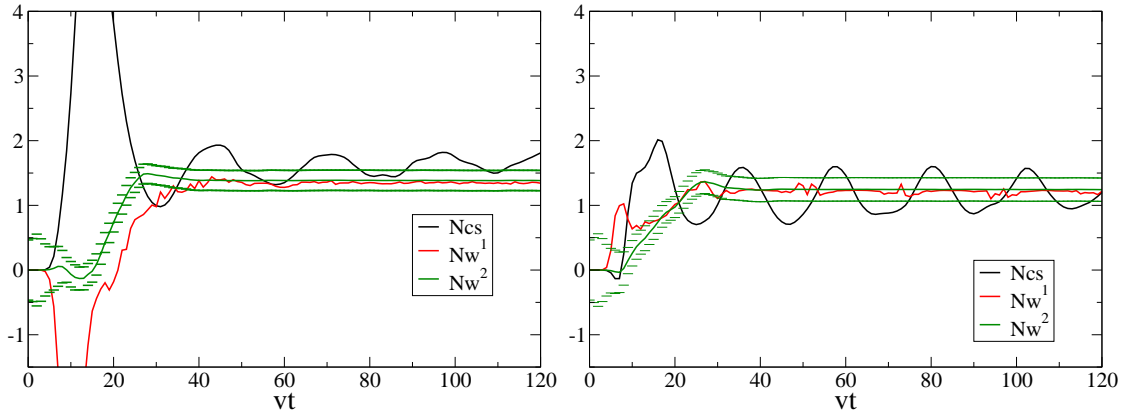

\begin{center}
\epsfig{file=./Pictures/averages_km5_2.eps,width=0.48\textwidth,clip}
\epsfig{file=./Pictures/averages_km5_realvevs.eps,width=0.48\textwidth,clip}
\caption{\label{fig:ave_complex} Chern-Simons number and Higgs winding numbers when averaged over an ensemble of $4\times 25$ configurations. The potential is CP-breaking with complex vevs (left) and real vevs (right). $\delta_{\rm C/P}=-105$. We have included the $1\sigma$ statistical error band for $N_{\rm w}^2$ for illustration. The errors on the other observables are similar.}
\end{center}
\end{figure}

In order to compute the expectation value of the baryon asymmetry, we now perform simulations of an ensemble of $4\times 25$ configuration (4 for the conjugates) and compute average Chern-Simons number and Higgs winding number, as shown in Fig.\ref{fig:ave_complex} for the complex-vev (left) and the real-vev (right) CP-breaking potentials. The error bars/band on $N_{\rm w}^2$ correspond to standard deviations as defined in Eq.~(\ref{eq:sigma}). 

For the complex-vev case, we see a large initial creation of Chern-Simons number, driven by the C-/P- violating term as the Higgs fields go through the transition. This initial growth is not matched initially by the winding numbers, but around time $t=20/v$, these also acquire a non-zero average value. Eventually by $t=30/v$, all three observables have settled to a common asymmetry, and nothing more happens. The only remaining effect is that Chern-Simon number oscillates in time with the Higgs fields, in effect the coefficient of the C-/P-breaking term. 

For the real-vev case, there is no initial bouncing of the Chern-Simons number, and together with the winding numbers there is a slow drift to a finite value, which within error bars matches the complex-vev result. This may seem surprising, but means that whether or not the late-time Higgs expectation value is real or complex is not very important for the asymmetry. Baryogenesis takes place in the initial very chaotic stage, where CP-violation is active through complex couplings irrespective of the reality of the vevs.

\begin{figure}
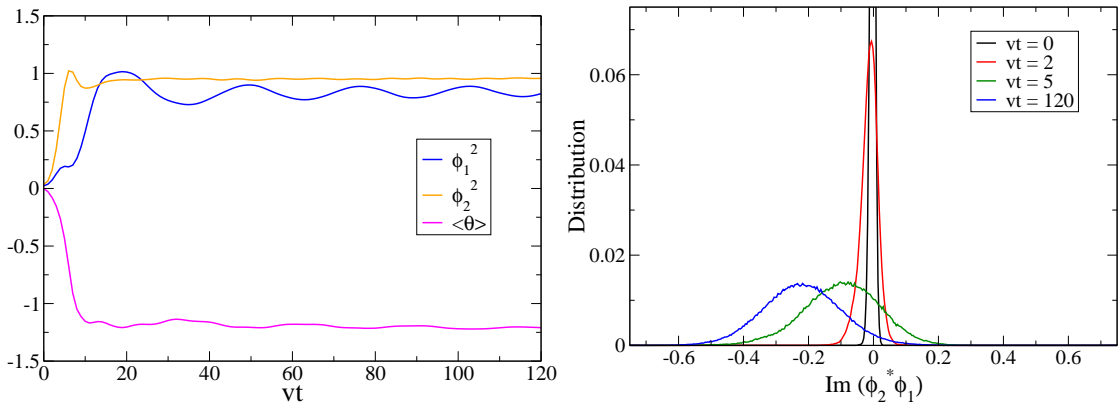

\begin{center}
\epsfig{file=./Pictures/averages_km5.eps,width=0.48\textwidth,clip}
\epsfig{file=./Pictures/hist_real5.eps,width=0.48\textwidth,clip}
  \caption{\label{fig:ave_angle_complex}  Left: The evolution of the normalized Higgs expectation values and the average complex angle $\theta$ for a CP-breaking, and complex-vev potential, and its evolution in time. 
Right: The distribution of the Higgs field imaginary part $|\phi_1||\phi_2|\sin\theta$ at the initial, intermediate and final times ($t=$0, 2, 5, 120).}
\end{center}
\end{figure}

P violation results from the C-/P-breaking term when the coefficient Im$(\phi_2^*\phi_1)=|\phi_1||\phi_2|\sin\theta$ is non-zero, effectively biasing Chern-Simon number. In Fig.~\ref{fig:ave_angle_complex} (left), we show the evolution of the (normalized) average Higgs fields and the relative angle $\theta$ for the complex-vev potential. All settle fairly early on, $t\simeq 10/v$, resulting in a non-zero P-violating coefficient. In Fig.~\ref{fig:ave_angle_complex} (right) we show the distribution of Im$(\phi_2^*\phi_1)(x)$ at the initial, final and two intermediate time-slices. We see how the initial condition is C-symmetric and strongly peaked, and then as the system evolves, the distribution flattens out and moves to a non-zero average value. 

\begin{figure}
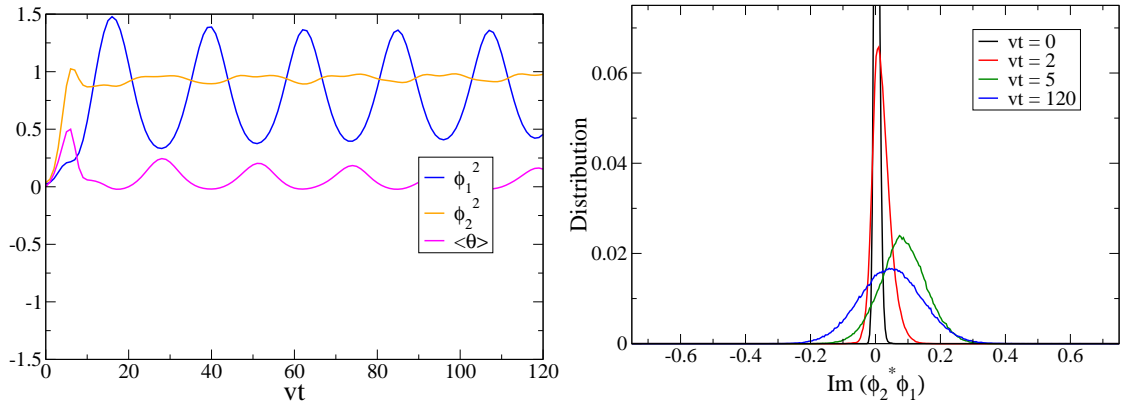

\begin{center}
  \centering
\epsfig{file=./Pictures/averages_km5_2_realvevs.eps,width=0.48\textwidth,clip}
\epsfig{file=./Pictures/hist_realvevs.eps,width=0.48\textwidth,clip}
  \caption{\label{fig:ave_angle_real} Left: The average complex angle $\theta$ for a CP-breaking, real-vev potential, and its evolution in time. 
Right: The distribution of the Higgs field imaginary part $|\phi_1||\phi_2|\sin\theta$ at the initial, intermediate and final times ($t=$0, 2, 5, 120).}
\end{center}
\end{figure}

For the real-vev case, CP-violation shows up as a nonzero average of the Higgs field angle Eq.~(\ref{eq:higgsangle}) at intermediate times. In Fig.~\ref{fig:ave_angle_real} (left) we show this average, and in Fig.~\ref{fig:ave_angle_real} (right) the distribution at different times. The distribution is again symmetric and peaked initially, but now during evolution the average moves away from zero, and then returns. The distribution again flattens. We also see that the late times average value is no longer exactly zero, showing that the minimum of the effective potential at finite temperature moves to complex expectation values, even though at tree level the vev is tuned to be real. In Fig.~\ref{fig:ave_angle_real} (left) we also shos the average fields, and again that the potential is rather shallow in the $\phi_1$-direction, so that at finite temperature the field oscillates with large amplitude. 

\begin{figure}
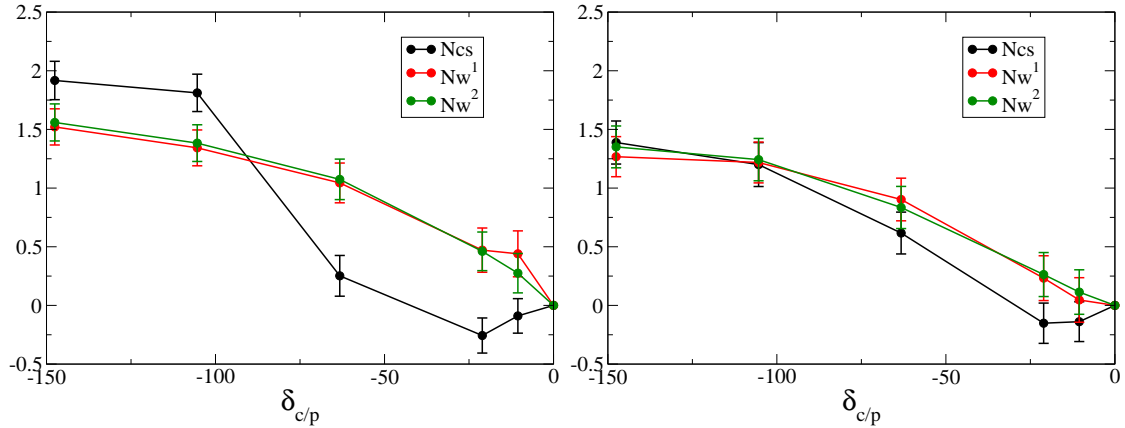

  \begin{center}
    \epsfig{file=./Pictures/kappadep_real5.eps,width=0.48\textwidth,clip}
    \epsfig{file=./Pictures/kappadep_realvevs.eps,width=0.48\textwidth,clip}
  \caption{\label{fig:delta_dep_real} The dependence of the baryon asymmetry on the strength of P-violation, for the CP-breaking, complex vev (left) and real vev (right) potentials.}
\end{center}
\end{figure}

The final baryon asymmetry depends on the dynamics of the system, and hence the precise form of the potential. For the present work we will not sweep through this rather large parameter space, but instead fix the potential, and calculate the dependence of the asymmetry on the parameter $\delta_{\rm C/P}$. The net Chern-Simons and winding numbers are shown for the complex- and real-vev cases in Fig.~\ref{fig:delta_dep_real}. We first remark that because at late times the Chern-Simons number is still oscillating, the cleaner observables are the winding numbers. These have a smooth, monotonic but non-linear behaviour as a function of $\delta_{\rm C/P}$, which goes very neatly through the origin as it must by construction.

Based on the result for $\overline{N}_{\rm w}^{1,2}$ at $\delta_{\rm C/P}=-21$, we conclude that the baryon asymmetry is given by
\ba
\frac{n_B}{n_\gamma}=-\delta_{\rm C/P}\times (2.8\pm 1.2)\times 10^{-6},
\ea
for the complex-vev potential and
\ba
\frac{n_B}{n_\gamma}=-\delta_{\rm C/P}\times (1.6\pm 1.2)\times 10^{-6},
\ea
for the real-vev potential. These two results are consistent with each other at the 1$\sigma$ level, and this agreement applies to the full $\delta_{\rm C/P}$-dependence in Fig.~\ref{fig:delta_dep_real}. Again, it seems to have little impact whether in an otherwise identical potential, the expectation value is rotated to a complex value or not. 

\section{Conclusion and outlook}
\label{sec:conclusion}

Using fully non-perturbative and out-of-equilibrium classical lattice simulations, we have computed the baryon asymmetry generated in a cold spinodal electroweak transition in the Two-Higgs-Doublet model. CP-violation appears as C-violation in the scalar potential, but only in combination with explicit P (and in our case, also C) breaking can a net Chern-Simon number density be created. In the full theory, this is present through the usual left-handed projector in the gauge-fermion interaction. In this bosonic model, this is replaced by the leading C-/P-breaking bosonic operator expected upon integrating out the fermions.

Our results show that an asymmetry is indeed created; it is large enough to be seen clearly in simulations, and is not suppressed by being a combination of two (small) effects, the CP-violation and the C-/P-violation. The observed cosmological baryon asymmetry of $\simeq 6\times 10^{-10}$ is reproduced for 
\ba
\delta_{\rm C/P}=-(2\textrm{ to }3)\times 10^{-4}.
\ea
The asymmetry is largely created in the initial spinodal roll-off, and the late time dynamics is mostly irrelevant. In particular, whether or not the Higgs vev is real or complex had little impact (if the potential is otherwise the same). 

In the potentials used here, the temperature after the transition is high enough that there is ``partial symmetry restoration'' in the lightest Higgs mode direction. This is partly because all the fermion degrees of freedom are not included dynamically and $g^*$ is much lower in the simulations. This restoration makes it possible for one of the Higgs fields to unwind, but since both winding numbers and Chern-Simons number are dynamically chained to each other at late times, this does not happen. At early times, Chern-Simons number and the winding numbers move independently.

The ``missing link'' is how to trigger the required spinodal transition and how the Universe came to be super-cooled way below the electroweak scale. This can be achieved through a further enlarged scalar sector, for instance by a gauge singlet, which may \cite{enqvist} or may not \cite{vantent} be taken to be the inflaton. Further examples of potentials, which may incorporate a first order transition as well, have been proposed in \cite{servant}.

The obvious next step is to compute the value of $\delta_{C/P}$, or in more generality, the functional form and coefficients of all C- and P- violating terms arising from integrating out the fermions. This can for instance be done using a gradient expansion as in \cite{smit4,schmidt6,salcedo6,salcedo8,CPV1}. The precise asymmetry generated will depend on the exact scalar potential, but we expect the $\delta_{C/P}$ dependence found here to persist for a given potential. A more detailed study of the space of scalar potentials, including cases with only one unstable initial field, would also be a natural extension of the present work. Finally, the inclusion of the U(1) gauge field, a dynamical scalar to trigger the spinodal transition and ultimately dynamical fermions would all be interesting extensions and improvements.

\acknowledgments
We would like to thank Markos Maniatis, Mikko Laine, Aleksi Vuorinen, Tomas Brauner and Jens Oluf Andersen for
helpful discussions. B.W. was supported by the Humboldt foundation through its Sofja
Kovalevskaja program. A.T. was supported by the Carlsberg Foundation.

\appendix

\section{Choice of Higgs potential parameters}
\label{app:parameters}

\begin{figure}
\begin{center}
\epsfig{file=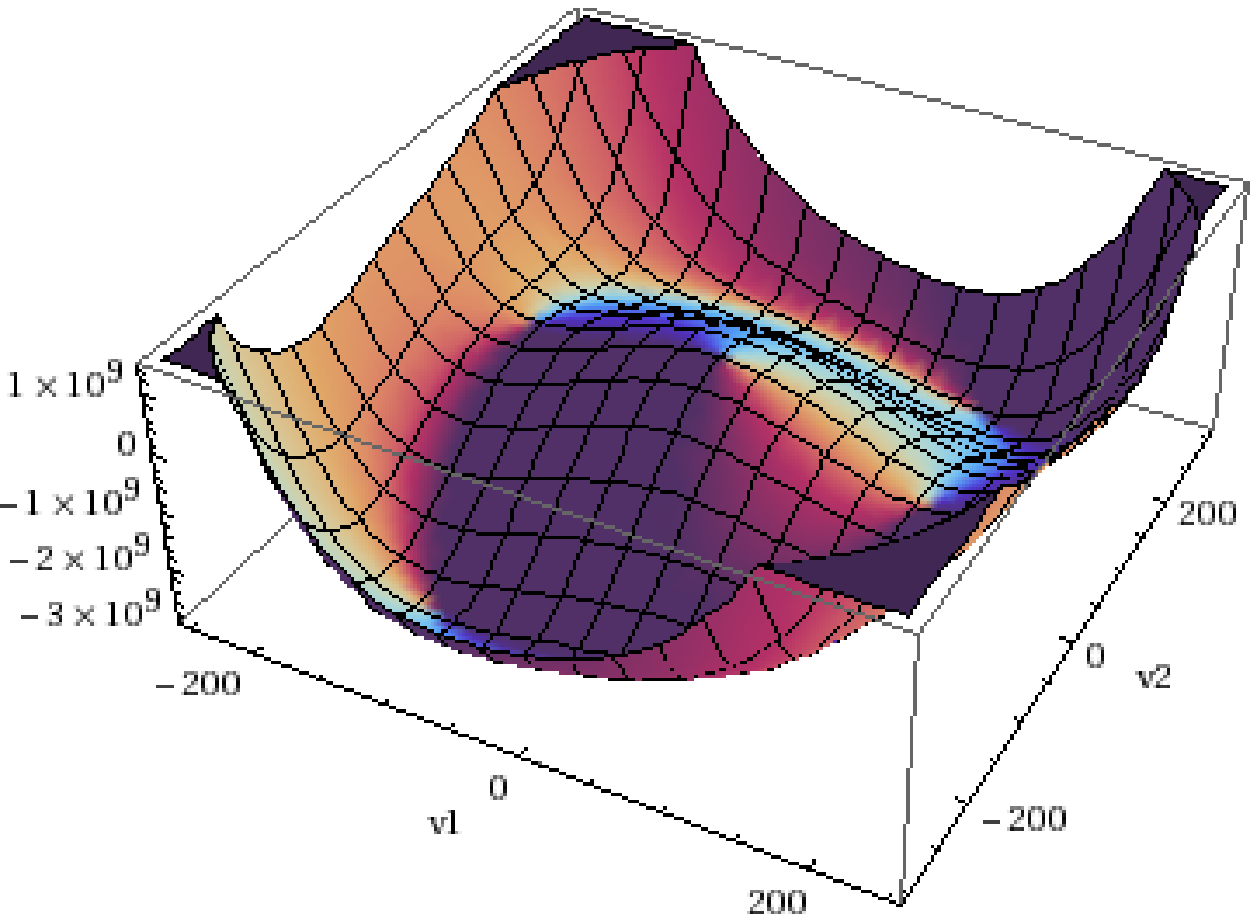,width=0.48\textwidth,clip}
\epsfig{file=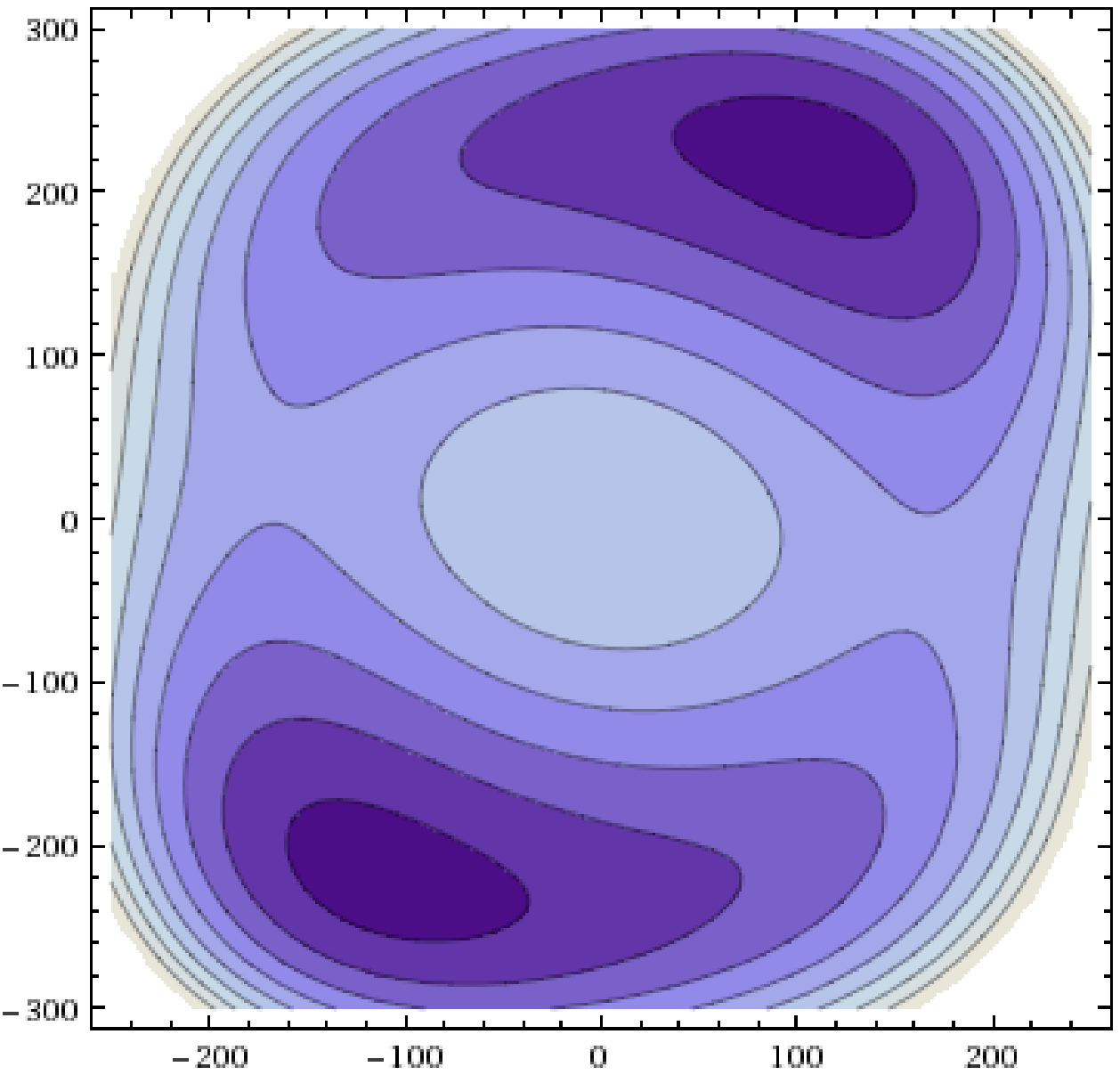,width=0.48\textwidth,clip}
  \caption{\label{fig:pot1} The potential for parameter set 1, CP-conserving, real vevs. Plotted in $v_{1,2}$-space at $\theta_2-\theta_1=0$.}
\end{center}
\end{figure}

\begin{figure}
\begin{center}
\epsfig{file=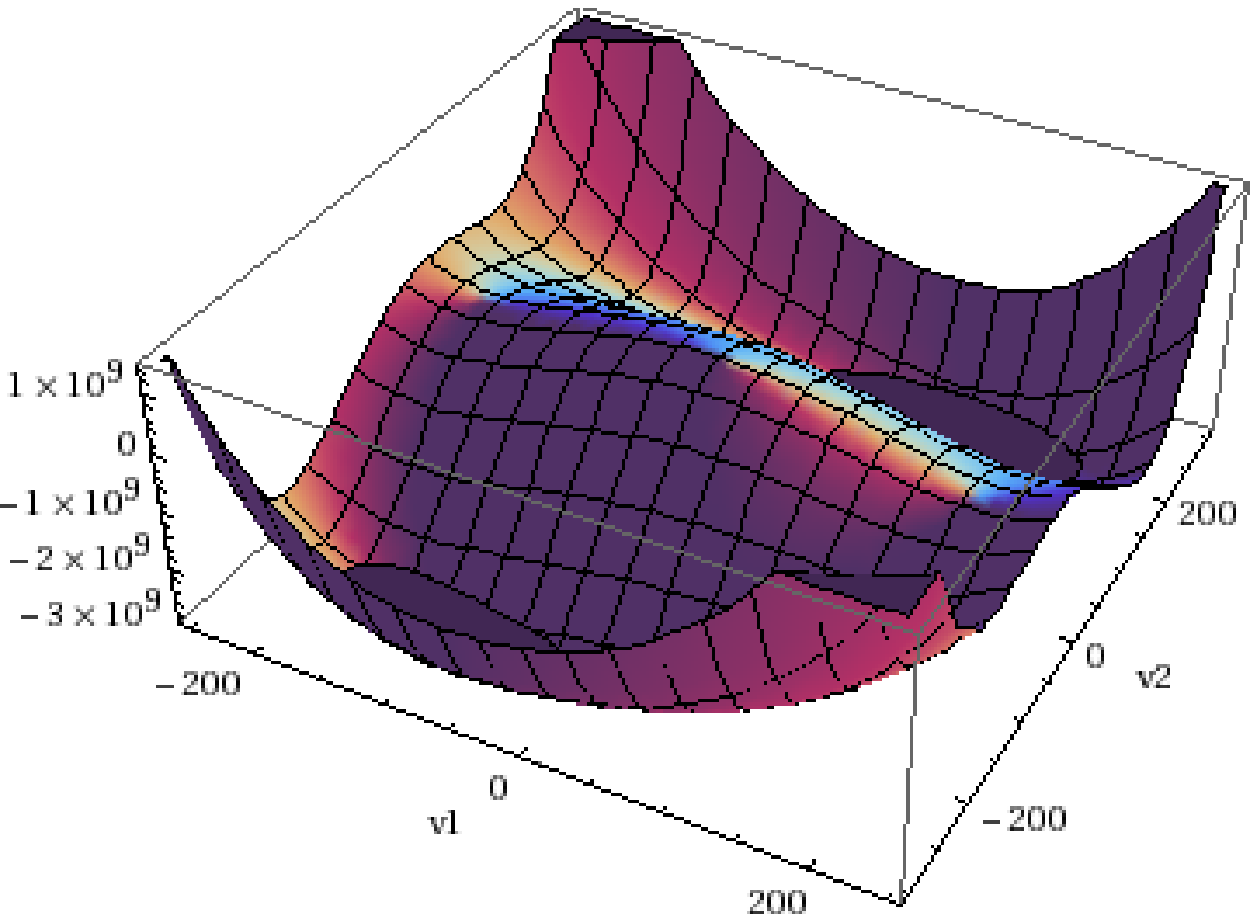,width=0.48\textwidth,clip}
\epsfig{file=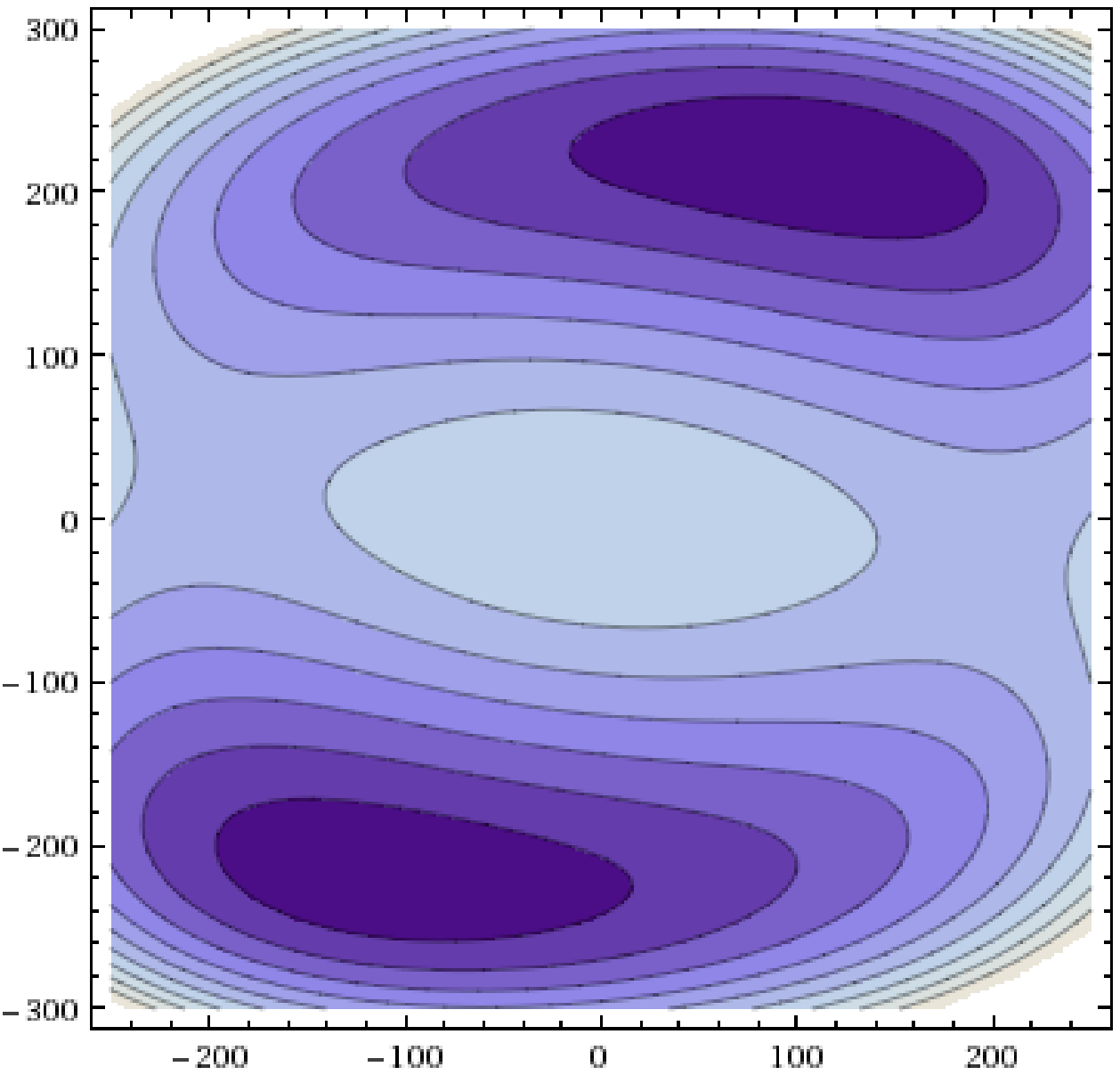,width=0.48\textwidth,clip}
  \caption{\label{fig:pot2} The potential for parameter sets 2 and 3, CP-breaking, real/complex vevs. Plotted in $v_{1,2}$-space, at $\theta_2-\theta_1=0$ and $1.39$, respectively.}
\end{center}
\end{figure}

We consider three different Higgs potentials, denoted as parameter sets 1, 2 and 3 respectively. They all obey our main criteria for the parameters $\lambda_{1,2,3,4,5}$ and the mass parameters $\mu^2_{11,22,12}$ that only $\mu_{12}^2$ and $\lambda_5$ can be complex. We impose the following constraints (See also \cite{2hdmreview,osland1,osland2}):
\begin{itemize}
\item The global minimum has $\phi_1 = (0,v_1/\sqrt{2}\, e^{i\theta_1})^T$, $\phi_2=(0,v_2/\sqrt{2}\,e^{i\theta_2})^T$, with $v_1/v_2=\tan\beta=2$ and $\sqrt{v_1^2+v_2^2}=246$ GeV, so that $v_1=220$ GeV, $v_2=110$ GeV. 
\item The mass eigenvalues around the global minimum are $(m_\pm,m_1,m_2,m_3)=(400,125,300,350)$ GeV.
\item In the CP-conserving case, both $\lambda_5$ and $\mu_{12}^2$ are real, and $\theta_{1}-\theta_2=0$.
\item In the real-vev case we choose $\theta_1-\theta_2=0$, in which case we have the constraint $\textrm{Im}\lambda_5v_1v_2=\textrm{Im}\mu_{12}^2$.
\item In the complex-vev case $\theta_{1}-\theta_2$ is free and we choose $\lambda_5$ to be real. 
\end{itemize}
This leaves two, three and three parameters, respectively to be fixed, and we use the sets (masses squared in GeV$^2$) :
\begin{itemize}
\item 1) CP conserving: (see Fig.~\ref{fig:pot1})
\ba
&\lambda_1=2.5575,\quad\lambda_2=1.5424,\quad\lambda_3=6.1176,\quad\lambda_4=-3.0570,\quad\lambda_5=-1.8177,\nonumber\\
&\mu_{11}^2=71125,\quad \mu_{22}^2=84713,\quad \mu_{12}^2=10000
\ea
The eigenvalues of the initial condition mass matrix are $(M^2_A,M^2_B)=(300.0^2,256.6^2)$ GeV$^2$ for which the lattice size of $(Lv)^3=(38.4)^3$ is easily large enough for sufficiently many spinodal modes to be present. 
\item 2) Real-vev: (see Fig.~\ref{fig:pot2})
\ba
&\lambda_1=0.86175,\quad\lambda_2=2.36749,\quad\lambda_3=5.7886,\quad\lambda_4=-3.5845,\quad\lambda_5=-1.2902+i 0.48231,\nonumber\\
&\mu_{11}^2=34673,\quad \mu_{22}^2=120680,\quad \mu_{12}^2=10000+i 11675
\ea
The eigenvalues of the initial condition mass matrix are $(M^2_A,M^2_B)=(351.2^2,178.9^2)$ GeV$^2$ for which the lattice volume is again large enough.
\item 3) Complex-vev: (see Fig.~\ref{fig:pot2})
\ba
&\lambda_1=0.86175,\quad\lambda_2=2.36749,\quad\lambda_3=5.7886,\quad\lambda_4=-3.5845,\quad\lambda_5=1.3774,\nonumber\\
&\mu_{11}^2=34673,\quad \mu_{22}^2=120680,\quad \mu_{12}^2=13268-i7763.
\ea
The global minimum has $\theta_2-\theta_1=1.39$. The third potential is simply a rotation of $\phi_2$ by an angle $\theta_2-\theta_1$ to make $\lambda_5$ real. So the potential and its minimum are equivalent, if the P-breaking coefficient were also rotated by the same complex phase. If $\delta_{\rm C/P}$ is kept constant, it is a different physical system.
\end{itemize}
We have for the three sets, respectively
\ba
\left(\frac{V(0,0)-V(v_1,v_2)}{v^4}\right)^{3/4}= \left(0.800,0.940,0.940\right)
\ea
which enters in the calculation of the final asymmetry Eq.(\ref{eq:fin_asym}).



\end{document}